\documentstyle[aps,epsfig,psfig,multicol]{revtex}      
\begin{document}
\title{
Dynamics of bright matter-wave solitons in inhomogeneous cigar-type
Bose-Einstein condensate
}
\author{F.Kh. Abdullaev$^{(a)}$\thanks{On leave from Phys. Techn. 
Institute, Tashkent, Uzbekistan}, A. Gammal$^{(b)}$, and 
Lauro Tomio$^{(a)}$}
\address{
$^{(a)}$ 
Instituto de F\'\i sica Te\'orica, Universidade Estadual Paulista,
Rua Pamplona, 145,  01405-900, S\~{a}o Paulo, Brazil\\
$^{(b)}$ 
Instituto de F\'{\i}sica, Universidade de S\~ao Paulo,
05315-970 S\~{a}o Paulo, Brazil \\
}
\date{\today}
\maketitle
\begin{abstract}
We discuss the possible observation of a new type of standing nonlinear 
atomic matter wave in the condensate: the nonlinear impurity mode.
It is investigated dynamical effects of a bright soliton in
Bose-Einstein-condensed (BEC) systems with local space variations
of the two-body atomic scattering length. 
A rich dynamics is observed in the interaction between the soliton 
and an inhomogeneity. Processes as trapping, reflection and transmission 
of the bright matter soliton due to the impurity are studied considering 
an analytical time-dependent variational approach and also by full numerical 
simulation. A condition is obtained for the collapse of the bright solitary 
wave in the quasi-one-dimensional BEC with attractive local 
inhomogeneity. 
\end{abstract}  
\begin{multicols}{2}

\section{Introduction}
Theoretical investigations of nonlinear collective excitations of 
matter waves, actually became a very interesting and relevant subject 
with the experimental observations of Bose-Einstein condensation in
vapors of alkali-metal atoms~\cite{Dalfovo}.
One of the interesting forms of localized waves of atomic matter 
are the {\it solitons} - moving stationary nonlinear wave packets.
Historically, it was first observed the {\it dark } solitons,
represented by objects with nontrivial topological properties. They
can exist in BEC systems for positive two-body scattering length 
($a_{s} > 0$), corresponding to repulsive interactions between atoms. 
They can be presented as {\it holes} in the background of the 
condensates~\cite{Den}.

The observation of solitons in BEC systems with negative 
two-body scattering length ($a_{s} < 0$) - so called {\it bright} 
solitons, are more complicated from the experimental point
of view. The difficulty is related to the instability of condensates
in two-dimensions (2D) and three-dimensions (3D), when the number of 
atoms $N$ exceeds the critical limit $N_{c}$, which is typically a 
number 
of the order of 1,500 atoms for Li$^7$.
This experimental limitation can be softened in case of  
quasi - one-dimensional (1D) geometry; i.e., for  BEC in highly 
anisotropic cigar-type 
traps. 
We should observe that, for a true 1D system, one does not expect 
the collapse of the system with increasing number of 
atoms~\cite{rup95,pla}. However, it happens that
a realistic 1D limit is not a true 1D system, with the density of
particles still increasing due to the strong restoring forces in the
perpendicular directions~\cite{perez,3D}.

The observation of bright matter-wave solitons in BEC with attractive 
interactions has been recently reported in Ref.~\cite{Hulet,Khaykovich}.
Theoretical models, explaining the observed phenomena, have
been considered in Refs.~\cite{Carr,Stoof}. It
should be noted that in principle it is possible to observe
the bright matter waves solitons in BEC with optical lattice.
The possibility of changing the sign of the effective dispersion
in  such lattices makes possible to generate bright solitons for  
repulsive condensates~\cite{Abd01,Trom}.
Thus, it represents an interesting possibility to control the 
dynamics of bright matter-wave solitons. For the control, in this
paper we suggest to use artificially induced inhomogeneities, 
by considering a variation in the {\it space} distribution of the 
atomic two-body scattering length. This variation
can be achieved by using optical methods, as detuned laser 
field~\cite{Fatemi}, or by means of Feshbach resonance
applying an external magnetic field~\cite{avaria}. 
Mathematically, this leads to the appearance of a coordinate dependent 
coefficient in the nonlinear term of the Gross-Pitaevskii (GP) equation.  
We will analyze this problem for a cigar-type condensate using a 
full numerical solution of the GP formalism and also a 
time dependent variational approach, which was successful in the
description of BEC dynamics~\cite{Stoof1,perez}. 
In spite of the well known difficulties of the variational approach it 
represents a good framework for a preliminary insight into the basic 
physical mechanism of the model.

We consider the local space variation of the atomic scattering 
length as related to a {\it nonlinear} impurity term in the
nonlinear Schr\"odinger equation (NLSE).
Previously, interaction of solitons with linear impurities have been 
performed in the sine-Gordon model~\cite{Kiv91,Good}
and in the 1D NLSE~\cite{For,Cao,Good2}.
It was shown that the interaction with the impurity mode leads to 
unexpected behaviors of the soliton. In particular, the soliton can 
be reflected by an {\it attractive} impurity. This happens due to a
resonant interaction between the soliton and an oscillating impurity 
mode.

For the description of the soliton dynamics,
the variational approach gives two coupled equations: for the soliton 
width and center-of-mass position. The oscillation frequencies of the 
width can be transferred resonantly to the oscillation frequencies of 
the center of
mass; and, as a result, the trapped soliton can escape from the 
inhomogeneity. 
We should also note that exists a solution representing a {\it 
standing} nonlinear atomic matter wave. The soliton, trapped by the 
nonlinear impurity, evolves to this solution.
In this work we estimate the values of the parameters for the observation 
of such a new type of nonlinear standing atomic matter wave.

The structure of the present work is as follows: 
In Sect. 2, we formulate the model for the quasi 1D BEC 
with nonlinear inhomogeneity and derivate the equations for the soliton
parameter using the time dependent variational approach. In 
Sect. 3 we present the analysis of the fixed points, frequencies of 
oscillations  for the width and center of mass.
Section 4 contains the numerical modeling of the system of ODE's and 
numerical simulations of the inhomogeneous GP equation, with a final
discussion on the nonlinear impurity mode.

\section{Formulation of the model}
The mean field equation for a Bose-Einstein-condensed system,
trapped by an harmonic potential, is given by the following
GP equation:
\begin{eqnarray}
{\rm i}\hbar\frac{\partial}{\partial t}\Psi(\vec{r},t)
&=&
\left[ -\frac{\hbar ^{2}}{2m}{\bf \nabla }^{2}+
\frac{m}{2}
\left(
\omega_{1}^{2}x_1^{2}+
\omega_{2}^{2}x_2^{2}+
\omega_{3}^{2}x_3^{2}
\right)
\right.\nonumber\\
&+& \left.
\frac{4\pi \hbar ^{2}\ a_s}{m}|\Psi(\vec{r},t)|^{2}\right]
\Psi(\vec{r},t)
\label{1}
\end{eqnarray}
where $a_{s}$ is the atomic scattering length,
$m$ is the mass of the atom, and the
wave-function $\Psi\equiv\Psi(\vec{r},t)$ is normalized to the
number of particles $N$.
In the present work, we assume a cylindrical highly anisotropic
trapped potential,
such that i.e.
$\omega_{1}=\omega_{2}\gg\omega_{3}$.
In this circumstance, one can approximate the field as \cite{perez}
\begin{equation}
\Psi(\vec{r},t) = R(x_1,x_2) Z(x_3,t).
\label{2}
\end{equation}
$R(x_1,x_2)\equiv R$ satisfies the 2D harmonic oscillator 
equation, that we assume is
in the ground-state, normalized to one:
\begin{equation}
\left[ -\frac{\hbar ^{2}}{2m}\left(
\frac{\partial^{2}}{\partial x_1^2}+
\frac{\partial^{2}}{\partial x_2^2}
\right)
+
\frac{m\omega_{1}^{2}}{2}(x_1^2+x_2^2)
\right]R = \hbar\omega_1 R ,
\label{3}
\end{equation}
\begin{equation}
R(x_1,x_2) =
\sqrt{\frac{m\omega_1}{\pi\hbar}}
{\rm e}^{-\frac{m\omega_1}{2\hbar}
(x_1^2+x_2^2)}
\label{4}.
\end{equation}
By substituting the Eqs.(\ref{2})-(\ref{4}) in Eq.(\ref{1})
and averaging over the transverse coordinates,
\begin{eqnarray}
{\rm i}\hbar\frac{\partial Z}{\partial t}
&=&
\left[ -\frac{\hbar ^{2}}{2m}
\frac{\partial^2}{\partial x_3^2}
+\frac{m \omega_{3}^{2}x_3^{2}}{2}
+ \hbar\omega_1 +
{2 a_s}{\hbar\omega_1}
|Z|^{2} \right] Z
\label{5}.
\end{eqnarray}
In this case, $Z\equiv Z(x_3,t)$ is normalized to the number of particles 
$N$.
Next, we redefine $Z$ and the variables, such that
\begin{eqnarray}
u(z,\tau)\equiv \sqrt{4|a_s|} Z(x_3,t)
{\rm e}^{{\rm i}\omega_1 t},\nonumber\\
\tau\equiv\frac{\omega_1 t}{2};\;\;\; 
z\equiv\sqrt{\frac{m\omega_1}{\hbar}}x_3;
\;\;\;\; \alpha \equiv -\left(\frac{\omega_3}{\omega_1}\right)^2
\label{6}.\end{eqnarray}
From now, we consider a simplified notation for the functions,
such that $u\equiv u(z,\tau)$, $u_\tau\equiv \partial u/\partial\tau$,
and $u_{zz}\equiv \partial^2 u/\partial\tau^2$. Thus, we
obtain the following 1D NLSE:
\begin{equation}
\mbox i u_{\tau} = - u_{zz} + \sigma |u|^{2}u -
\alpha z^{2}u
\label{GP},
\end{equation}
where $\sigma$ is the signal of the two-body scattering length.
The equation (\ref{GP}), for $\alpha=0$ ($\omega_3\to 0$) and
$\sigma=-1$,
has the solitonic solution
\begin{equation}
u^{(s)} = \sqrt{2}\;A\;\mbox{sech}\left[A(z - v\tau)\right]
{\rm e}^{\mbox i
\left[\displaystyle
\frac{v z}{2} +
A^2\tau-\frac{v^2\tau}{4}
\right]},
\label{8}\end{equation}
where $A$ is a constant and $v$ is the soliton velocity.
In the following, let us consider
the interesting case of local space variation of a negative atomic
scattering length, such that Eq.~(\ref{GP}) is replaced by
\begin{equation}
{\rm i}u_{\tau} = - u_{zz} - \alpha z^{2}u - 
(1 + \epsilon f(z))|u|^{2}u.
\label{9}
\end{equation}
$\epsilon>0$ ($\epsilon<0$) refers to negative (positive) variation of 
the scattering length.
The corresponding 1D Hamiltonian energy is given by
\begin{equation}
\langle H\rangle = \frac{1}{n_0}
\int_{-\infty}^{\infty}dz\left[|u_{z}|^2 - \alpha z^2|u|^2-
\frac{\left(1+ \epsilon f(z)\right)}{2}|u|^4\right],
\label{10} \end{equation} 
where $n_0$ is the normalization of $u$, 
related to the number of particles $N$ and the scattering 
length $a_s$: 
\begin{equation}
n_0 = 4N|a_s|\sqrt{\frac{m\omega_1}{\hbar}}
.\label{11}\end{equation}

We should observe that, in the realistic case one has a quasi-1D
(cigar-like) trap. In Ref.~\cite{huletsol}, the authors have 
consider the formation and propagation of matter wave solitons, 
using a gas of $^7$Li atoms, in a quasi-1D trap. The frequencies 
used in their trap are  $\omega_\perp = \omega_1 = 2\pi\times$ 625 Hz, and 
$\omega_L=\omega_3= 2\pi\times$ 3.2 Hz, with the scattering length tuned 
to $a_s=-3a_0$ ($a_0$ is the Bohr radius). The ratio of such frequencies
gives $-\alpha=(\omega_3/\omega_1)^2=2.6\times 10^{-5}$.
In this case, as shown in Ref.~\cite{3D}, the maximum critical number 
$n_{0,max}\approx 2.70$ is a constant that does not depend on $a_s$.
The realistic maximum number of atoms $N_c$, will be related 
to $a_s$ and the oscillator length:
$4 N_c|a_s|\sqrt{{m\omega_1}/{\hbar}}\approx 2.70$.

To study the dynamics of the perturbed soliton we use 
the following trial function~\cite{Anderson}:
\begin{equation}
u = A\;\mbox{sech}
\left(\frac{z-\zeta}{a}\right)\;
{\rm e}^{\mbox i
\left[\phi + w\left(z-\zeta\right)+b\left(z - \zeta\right)^{2}
\right]}
\label{12},
\end{equation}
where $A$, $a$, $\zeta$, $\phi$, $w$, and $b$, are time-dependent
variational parameters. In this case $u$ is normalized to
$n_0=2aA^2$.
To derive the equations for the time-dependent parameters of the
soliton, we first obtain the averaged Lagrangian
\begin{equation}
\bar{L}(\tau) = \int {\cal L}(z,\tau)dz
\label{13},
\end{equation}
with
\begin{eqnarray}
{\cal L}(z,\tau) &=& \frac{\rm i}{2}(u_{\tau}u^{\ast} - 
u_\tau^{\ast}u)-|u_{z}|^{2} + \alpha z^2 |u|^{2}\nonumber\\
&+& \frac{1}{2}[1+\epsilon f(z)]|u|^{4}.
\label{14}
\end{eqnarray}
The equations for the soliton parameters are derived 
from the Lagrangian $\bar{L}$, by using the corresponding 
Euler-Lagrange equations. So, $\bar{L}$ is given by
{\small
\begin{eqnarray}
\bar{L} &=& -n_0\left[\phi_{\tau} - w\zeta_{\tau} +
\frac{\pi^2}{12}a^{2}b_{\tau}\right]
-\frac{n_0}{3a^2} - n_0 w^2 
-\frac{\pi^2 n_0\;a^2 b^2}{3}
\nonumber\\ &+&
\frac{n_0^2}{6a} + \epsilon\frac{n_0^2}{8a^2}F(a,\zeta) +
\alpha\; n_0 \left(\zeta^{2}+\frac{\pi^2}{12}a^2
\right)
\label{15},
\end{eqnarray}
}
where
\begin{equation}
F(a,\zeta)\equiv
\int_{-\infty}^{\infty}dz\frac{f(z)}{\cosh^{4}(\frac{z-\zeta}{a})}.
\label{16}
\end{equation}
We also obtain the coupled equations for $a$ and $\zeta$:
\begin{eqnarray}
a_{\tau\tau}& =&\frac{16}{\pi^{2}a^{3}} - \frac{4\;n_{0}}{\pi^2 a^2}-
\epsilon\frac{3\;n_0}{\pi^2 a^2}\left[2 \frac{F}{a} -
\frac{\partial F}{\partial a}\right] +
4\alpha a,\nonumber \\
\zeta_{\tau\tau}&=&4\alpha\zeta  +
\epsilon\;\frac{n_{0}}{4a^{2}}
\frac{\partial F}{\partial\zeta}.
\label{17}
\end{eqnarray}
When $a$ is constant we have the well known description of the soliton
center as the unit mass particle in an anharmonic potential
$U(\zeta) = -2\epsilon\;A^4 \mbox{sech}^{4}(A\zeta)$ \cite{Abd1}.
From this point of view, for $\epsilon < 0$, a slowly moving soliton
$(v_{s} < v_{c}= 2\sqrt{|\epsilon|})$ can  be trapped by the impurity.
Considering the impurity with $\epsilon < 0 $, the soliton can pass
through the impurity or reflect, depending on its velocity.
The numerical simulations of the system (\ref{17}) and the GP equation
(see section 4) shows that the dynamics is much more complicated. In
particular we can observe the soliton reflecting from an attractive
impurity even when $v_{s} < v_{c}$.

In order to have a more general formulation of the model, in the
present section we have considered a non-zero external potential, 
parametrized by $\alpha$. One could also explore the behavior 
of the soliton, by considering a more general time-dependent form of the 
external potential, as studied in Ref.~\cite{nogami}.
However, in the present work our main motivation is the 
propagation of matter wave solitons, in a 1D cigar-like 
trap~\cite{huletsol}, such that we will assume $\alpha=0$ in the next 
sections.

\section{Dynamics of bright solitons under two kind 
of inhomogeneities}
\subsection{Point-like non-linear impurity}
Let us  first consider an inhomogeneity given by a delta type 
($f(z)=\delta(z)$), and look for the fixed 
points of the system of Eqs.~(\ref{17}). In this case,
\begin{equation}\label{18}
\left\{
\begin{array}{l}
F(a,\zeta)=\mbox{sech}^{4}\left({\zeta}/{a}\right),\\
{\partial F}/{\partial\zeta}
=-({1}/{a})\tanh\left({\zeta}/{a}\right)
\mbox{sech}^{4}\left({\zeta}/{a}\right),\\
{\partial F}/{\partial a}
=({\zeta}/{a^2})\tanh\left({\zeta}/{a}\right)
\mbox{sech}^{4}\left({\zeta}/{a}\right).
\end{array}
\right.
\end{equation}
The fixed point for the soliton center is given by  $\zeta =0$. This
corresponds to the case of an atomic matter soliton
trapped by the local variation of the two-body scattering length.
In case the local variation corresponds to a positive
scattering length ($\epsilon < 0$), we should observe the
soliton being reflected by the inhomogeneity. Then, 
the stationary width $a_{c}$ can be defined by
\begin{equation}
a_{c} = \frac{8-3\epsilon \;n_{0}}{2\;n_{0}}.
\label{19}
\end{equation}

Expanding the solution near $a_{c}$ we obtain the frequencies of
small oscillations for the width $a$ and for the center-of-mass  
$\zeta$ of the soliton, localized by the impurity. The square of
such frequencies are, respectively, given by

\begin{eqnarray}
\omega_{a}^{2} &=& \frac{4\;n_0}{\pi^2}\left(
\frac{2\; n_{0}}{8-3\;n_{0}\;\epsilon}\right)^{3}
\nonumber\\
\omega_{\zeta}^2 &=&
\epsilon n_0 \left(\frac{2\;n_{0}}{8-3\;n_0\;\epsilon}\right)^4.
\label{20}
\end{eqnarray}

In the variational approach for the soliton  interacting with the impurity
we have the interaction of the oscillating internal degree of freedom
(the width) with the soliton center. As can be seen from Eqs.~(\ref{20}),
the frequencies of the oscillations can match for a certain value of 
$\epsilon$, with energy transfer between the two modes.
So, as a result of the reflection from attractive 
inhomogeneity in BEC, a depinning of the soliton can occur.

\subsection{Interface between two BEC media}
Now, we consider another interesting case of an interface between
two media, such that at $z=0$ we have a sudden change in the
two-body scattering length. The size of the inhomogeneity is
given by $\epsilon$ in Eq.(\ref{9}), where $f(z)=\theta(z)$.
In this case, we have
\begin{equation}
\left\{
\begin{array}{l}
F(a,\zeta)= a\left[{2}/{3}+
\mbox{tanh}\left({\zeta}/{a}\right)
-\frac{1}{3}\mbox{tanh}^3\left({\zeta}/{a}\right)\right],\\
{\partial F}/{\partial a}={F}/{a}
-({\zeta}/{a})\mbox{sech}^{4}\left({\zeta}/{a}\right),\\
{\partial F}/{\partial\zeta}
=\mbox{sech}^{4}\left({\zeta}/{a}\right).
\end{array}
\right.
\label{21}
\end{equation}
And, from (\ref{17}) with $\alpha=0$, we obtain the coupled equations:
\begin{eqnarray}\label{22}
a_{\tau\tau}& =& 
\frac{16}{\pi^{2}a^{3}}
-(\epsilon+2)\frac{2\;n_0}{\pi^2 a^2}
-\epsilon\frac{3\;n_0}{\pi^2 a^2}
\left[\mbox{tanh}\left(\frac{\zeta}{a}\right)
\right. \nonumber\\
&-&\left.
\frac{1}{3}\mbox{tanh}^3\left(\frac{\zeta}{a}\right)+
\left(\frac{\zeta}{a}\right)
\mbox{sech}^{4}\left(\frac{\zeta}{a}\right)\right],
\label{23}\\
\zeta_{\tau\tau}&=&
\epsilon\;\frac{n_{0}}{4a^{2}}
\mbox{sech}^{4}\left(\frac{\zeta}{a}\right)
\label{24}.
\end{eqnarray}
There is no fixed point. 
At the interface, the value of the width is reduced, 
\begin{equation}
a_{int} = \frac{8}{n_0(\epsilon+2)},
\end{equation}
and the frequency of oscillation of the pulse
width is
\begin{equation}
\omega_{a} = \frac{n_0^2(\epsilon+2)^2}{16\pi}.
\end{equation}
For a constant value of $a$, 
from (\ref{24}), we obtain
\begin{eqnarray}
\zeta_{\tau}^2=
2\epsilon\;\frac{n_{0}}{4a^{2}}F(a,\zeta)\le
\epsilon\;\frac{2\;n_{0}}{3a}.
\label{25}
\end{eqnarray}
When $a=4/n_{0} $ the system  (\ref{22}) reduces to the single equation 
for $\zeta$ describing the motion of the effective particle, for
optical beam crossing the interface of two nonlinear Kerr media  
considered in \cite{Nes,Ace}. 
If the velocity  exceeds the critical value, the soliton pass
trough the inhomogeneity. An interesting effect, predicted in 
Ref.~\cite{Ace}, can occur when the soliton cross the interface, namely, 
the possibility of soliton splitting. The soliton is the solution in the 
first medium. 
In the second medium it can be considered as the initial 
wavepacket deviating from the solitonic solution, for this media.
Applying the approach developed in Ref.~\cite{Man},
such initial condition will decay on few solitons plus 
radiation~\cite{Ace}.
The number of generated solitons is equal to 
\begin{equation}
n_{sol} = {\cal I}\left[
\frac{1}{\sqrt{2}\pi}\int_{-\infty}^{\infty}|u_0|dz + 
\frac{1}{2}
\right],
\label{26}\end{equation}
where $u_0$ is the initial solution for the second medium, and 
${\cal I}[...]$ stands for {\it integer part of} $[...]$.
Thus, if we have a jump in the scattering length given by $\Delta a_{s} = 
a_{s2} -a_{s1}$, then the number of generated solitons in the
second part of the BEC is equal to 
$ n_{sol} = {\cal I}\left[\sqrt{a_{s2}/a_{s1}}+1/2\right]$,
where $(1+\epsilon) = a_{s2}/a_{s1}$ is the ratio of 
the atomic scattering lengths.
For example, for the ratio 9/4$<a_{s2}/a_{s1}<$25/4 
(or 1.25$<\epsilon<$5.25) we obtain two solitons in the right-hand-side 
medium. To obtain $n_{sol}$, we need $\epsilon$ such that  
$n_{sol}(n_{sol}-1) < (\epsilon+3/4) < n_{sol}(n_{sol}+1)$.

\section{Numerical simulations and general discussion}

Our approach for pulses deviating from the exact soliton solution
is interesting from the experimental point of view, considering the
difficulty in producing exact solitonic solutions.
Is of particular interest the non-trivial case of nonlinear 
Dirac-delta impurity ($f(z)=\delta(z)$), where we made detailed 
comparison between the variational and full numerical solution of the
GP equation. 
In Fig. 1, we are comparing the variation results with the numerical ones,
for fixed point of the width, given by $a$ (top frame);
for the frequency of oscillations of the width, $\omega_a$ (middle 
frame); and for the frequency of oscillations of the center-of-mass, 
$\omega_\zeta$, trapped by the inhomogeneity (bottom frame).
By using the variational approach, we observe that the width goes to 
zero and the frequencies are singular when 
\begin{equation}
\epsilon = \epsilon_c = \frac{8}{3\;n_0}
\label{27}.
\end{equation}
Here, it is interesting to observe that we have two critical numbers 
that are related: one of the critical number $n_{0,max}\approx 2.7$ 
comes from the quasi-1D limit of a 3D calculation~\cite{3D}; another, 
is the maximum amplitude of the delta-like impurity that we have just 
introduced, given by (\ref{27}).
Considering both restrictions, we have that the smaller value
of $\epsilon_c$ is about one.

The plots in Fig. 1 are valid for any value of $n_0$ (where the 
maximum is about 8/3, according to \cite{3D}), because the width and
the frequencies were rescaled, such that
$\epsilon^\prime\equiv({n_0}/{4}) \epsilon$ (implying that
$\epsilon^\prime_c=2/3$), and
\begin{eqnarray}
{a}^\prime&\equiv& \frac{n_0}{4} a =
\left(1-\frac{3}{2}\epsilon^\prime\right),\nonumber\\
{\omega}^\prime_a &\equiv& \left(\frac{4}{n_0}\right)^2 \omega_a = 
\frac{4/\pi}{\sqrt{\left(1-{3}\epsilon^\prime/2\right)^{3}}},\nonumber\\
{\omega}^\prime_\xi&\equiv& \left(\frac{4}{n_0}\right)^2 \omega_\xi = 
\frac{2\sqrt{\epsilon^\prime}}{\left(1-{3}\epsilon^\prime/2\right)^{2}}.
\nonumber
\end{eqnarray}
As shown in Fig. 1, the variational results are supported by the full 
numerical calculation.
The singularity occurs when the contributions coming from the
inhomogeneity and nonlinearity are equal to the contribution from the
quantum pressure, as seen from Eq.~(\ref{17}).
When $\epsilon\ge \epsilon_c$, occurs the collapse of solitary wave.
So, we can observe the collapse of a 1D soliton on the attractive 
nonlinear impurity. 
This possibility can be obtained following a 
dimensional analysis in the 1D Hamiltonian given in Eq.~(\ref{10}).
The behavior of the field at small
widths is  $u \sim 1/L^{1/2}$. Taking into account that
$\delta(z) \sim 1/L \sim |u|^2$, we can conclude that the contribution of 
the potential energy due to the impurity is $\sim |u|^6$. For positive $\epsilon$, this term 
on the impurity  exceeds the quantum pressure and leads to the collapse of 
the soliton. In real situations, the collapse will be arrested on the 
final stage of the evolution, when the soliton width become of the order 
of the inhomogeneity scale. Then, the delta-function 
approximation for the impurity will break up.
\begin{figure}
\setlength{\epsfxsize}{1.0\hsize}
\centerline{\epsfbox{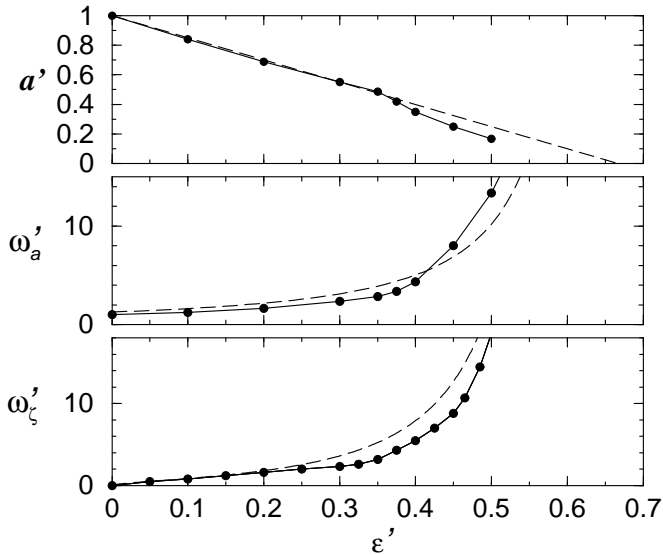}}
\caption{
The width ${a}^\prime\equiv{a (n_0/4)}$, the frequency of the 
width oscillations ${\omega}^\prime_a\equiv\omega_a (4/n_0)^2$, 
and the frequency of the soliton center oscillations 
${\omega}^\prime_\zeta\equiv\omega_\zeta (4/n_0)^2$, versus
the strength of the nonlinear delta-like impurity 
$\epsilon^\prime\equiv\epsilon (n_0/4)$. 
Solid line corresponds to full numerical solution of the GP equation, 
and dotted line to the corresponding variational approach.
All the quantities are in dimensionless units, as explained in the
text. 
}
\end{figure}
\begin{figure}
\setlength{\epsfxsize}{0.9\hsize}
\centerline{\epsfbox{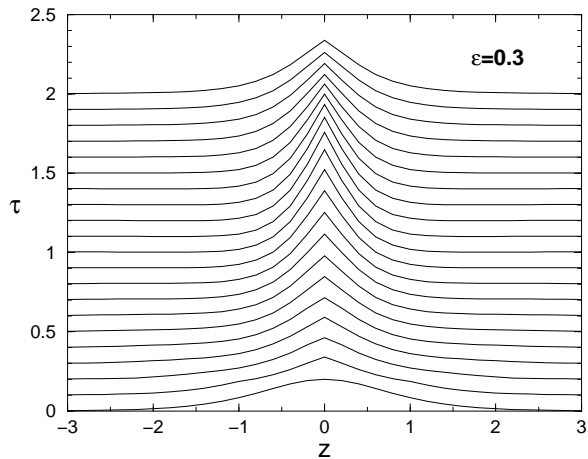}}
\caption{Density profile evolution for a fixed value of the 
amplitude of the delta-like impurity, $\epsilon=0.3$, in a 
projected 3D plot. At $z=0$, we observe the oscillation of the
amplitude, starting from the normal one and going to 
the nonlinear localized one. Each line represents a fixed value
of $\tau$.
The plot is shown for a moderately small value of $\epsilon$,
in order to avoid crossing of the lines. For larger values of 
$\epsilon$ we can also observe the emission of radiation.
$\tau$ and $z$ are dimensionless quantities, as given in the text.
}
\end{figure}

In Fig. 2, we show numerical simulations of the wave profile. We note 
that, after strong emission of radiation, it evolves into the so-called 
{\it nonlinear localized mode}.
The nonlinear localized mode represents an {\it exact} solution
of GP equation with nonlinear impurity (\ref{9}) and it is the nonlinear
{\it standing} atomic matter wave. The solution is given by
\begin{equation}
u^{(ni)} = \sqrt{2}a\mbox{sech}\left[a|z| + \beta
\right]e^{{\rm i}a^2\tau},
\label{uni}
\end{equation}
where
\begin{equation}
\beta\equiv\beta(\epsilon,a)\equiv \mbox{sign}(\epsilon)
\ln\left({2|\epsilon|a}+{\sqrt{1+4\epsilon^2a^2}}\right)^{1/2}
\label{beta}
\end{equation}

This solution can be obtained by using the solution of the homogeneous
equation with the requirement of the field continuity at the 
inhomogeneity and satisfying the jump condition in the first
derivative \cite{Sukh}.
The normalization $N^{(ni)}$, related to the umber of atoms for this 
solution is 
\begin{eqnarray}
N^{(ni)} &=& 4a[1 - \epsilon\gamma],\;\;\;
\gamma \equiv \gamma(\epsilon,a) \nonumber\\
\gamma &=&
\frac{\sqrt{1 + 4\epsilon^2 a^2 }-1}{2\epsilon^2 \;a}
=\frac{2\;a}{\sqrt{1+4\epsilon^2 a^2 }+1}
.\label{Nni}\end{eqnarray}

For small amplitude (or small impurity strength $|\epsilon|$)  we obtain
\begin{equation}
N^{(ni)} \approx 4a (1 - \epsilon \;a).
\end{equation}
At large amplitudes, we have
$N^{(ni)} \rightarrow 2/\epsilon$ for $\epsilon>0$;
and $N^{(ni)} \rightarrow 8a$ for $\epsilon<0$.
Note that for $\epsilon<0$ we have a solution with two bumps
structure for the nonlinear localized mode. As shown in \cite{Sukh}, this
mode is unstable. Here, we have considered only the case 
$\epsilon>0$.

In order to verify the stability of the solutions, one can study
the behavior of the second time derivative of the mean-square radius, as 
in Refs.~\cite{pla,Sukh,tsurumi}.
To obtain the second time derivative of the mean-square radius,
we use the Virial approach, with $H=-\partial_{zz}+V$ and
$V\equiv -(1+\epsilon\delta(z))|u|^2$:
\begin{eqnarray}
\langle z^2\rangle_{\tau\tau} &=& 
4\langle[H,z\partial_z]\rangle=
8\langle (-\partial_{zz})\rangle-4\langle 
zV_z\rangle,\nonumber\\
\langle zV_z\rangle&=&
=-\frac{1}{2}\langle V\rangle 
+\frac{\epsilon}{2n_0}|u_0|^4,\nonumber\\
\langle z^2\rangle_{\tau\tau} &=& 
\frac{1}{n_0}\int dz (8|u_z|^2-2|u|^4) 
-\frac{4\epsilon}{n_0}|u_0|^4.
\end{eqnarray}
For the system to collapse we need 
$\langle z^2\rangle_{\tau\tau} < 0$; implying that
\begin{equation}
\epsilon >  
\frac{1}{2|u(0)|^4}\int(4|u_{z}|^2 - |u|^4)dz
.\label{eps}\end{equation}
Using our solitonic ansatz, when $a\to 0$, we reach the
critical limit, $\epsilon_c = 8/(3n_0)$
that was obtained before.
\begin{figure}
\setlength{\epsfxsize}{1.0\hsize}
\centerline{\epsfbox{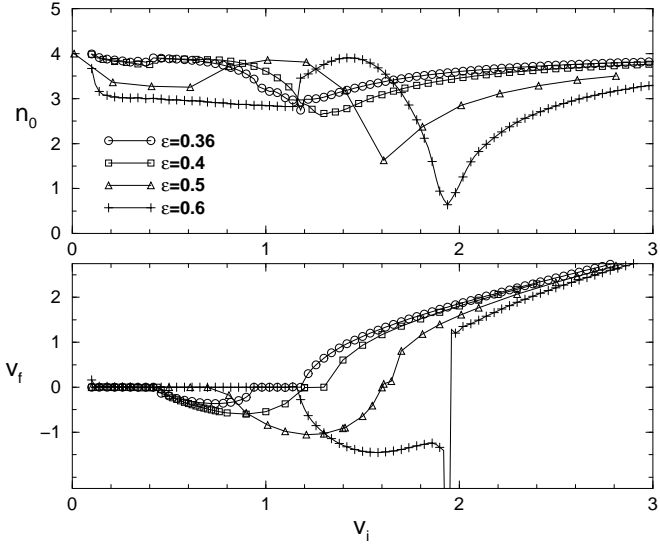}}
\caption{Numerical simulations of the full GP equation, showing 
the dependence of $n_0$, related to the number of atoms $N$ 
 (top frame), and final velocity $v_f$ (bottom frame), with respect to 
the initial velocity $v_i$. The results of both frames are shown for 
different values of $\epsilon$, as indicated inside the top frame. 
As shown, $n_{0}=4$ was considered the initial value (for $v_i=0$) 
of $n_0$. All the quantities are dimensionless.} 
\end{figure}

We have also investigated the dynamics of the matter soliton interacting
with inhomogeneity, studying different regimes of propagation
for several values of $\epsilon$. In Fig. 3, we present the results of
numerical simulations for the final velocity ($v_f$) versus the
initial velocity ($v_i$) of the soliton, considering different strengths
$\epsilon$ for the inhomogeneity.
In the present and next numerical approaches, we have considered
$n_0=4$, for the general 1D NLSE with nonlinear impurity,
as one can easily rescale the obtained data to a value of
$n_0$ smaller than 2.7, in agreement with the quasi-1D 
results.

As observed in Fig. 3, exists a region for the velocities where the {\it
attractive} nonlinear impurity reflects the soliton. In the
model involving the constant width approximation, this region
corresponds to the trapped soliton. The numerical results show
that always exists one window corresponding to the reflection of
the soliton. By increasing $\epsilon$, this window is shifted to
larger initial velocities.  From the variation of the number $N$, with
respect to the initial velocity $v_i$ (top frame of Fig.3), one
can also observe strong wave emissions by soliton, when $\epsilon$ 
increases and tends to the critical value (see also \cite{Sukh}). 
This picture reminds the picture of the collapse in 2D BEC with 
attractive interaction. We note that, by considering the interaction of
sine-Gordon kink with attractive defect, many windows were found,
corresponding to a resonance with local mode (see \cite{Kiv91}).

In order to compare with the results given in the lower frame of Fig.3,
we present in Fig. 4, for a fixed value of $\epsilon=0.4$, full
numerical calculation of the time evolution of the center-of-mass position, 
considering different values of the initial velocity. 
It is clearly seen the repulsive trajectories for $v_i$ corresponding to 
the case that the effective particle is trapped.

\begin{figure}
\setlength{\epsfxsize}{1.0\hsize}
\centerline{\epsfbox{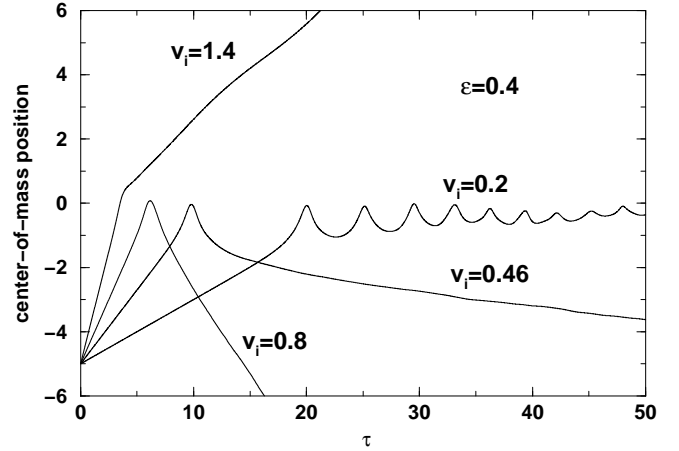}}
\caption{
For a fixed value of $\epsilon=0.4$, it is shown the time 
evolution of the center-of-mass position, considering different values
of the initial velocity, as given inside the frame.
The results were obtained by using full numerical solution of the
GP equation, considering $n_{0} =4$. All the quantities are in 
dimensionless units.
}
\end{figure}

By considering numerical simulations of the variational equations 
(\ref{18}) we found, qualitatively, the same behavior as the one 
observed in the full solution of the GP equation (see Figs. 3 and 5). 
In Fig. 5, we show the behavior of the final velocity as a function of 
the initial velocity, for different values of $\epsilon$.
In particular, the variational equations also show the existence of one 
window when the soliton is reflected by the impurity. When $\epsilon$ 
increases the window moves in the right direction, reducing the 
width.

\begin{figure}
\setlength{\epsfxsize}{1.0\hsize}
\centerline{\epsfbox{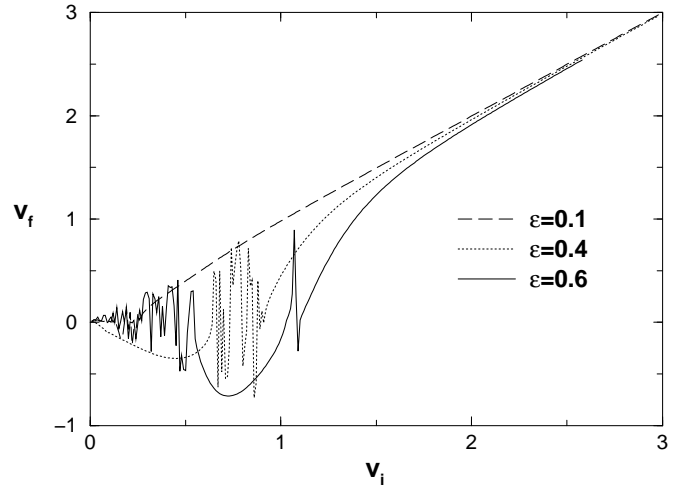}}
\caption{
Final velocity $v_f$ versus the initial velocity $v_i$, for 
different values of $\epsilon$, as indicated inside the frame, using  
$n_{0} =4$. The results were obtained from numerical simulations of 
the variational equations given in Eq.~(\ref{17}). All the quantities
are in dimensionless units.
}
\end{figure}

In distinction, on the GP equation, the variational equations shows a 
more complicated (probably chaotic) dynamics near the points where the 
regime of reflection starts or finishes. In this region, we also can see 
the more rare events with the transmission of soliton through the
impurity. This observation resembles the phenomena observed at the
interaction of the sine-Gordon equation kink with a local defect.
The system of ODE's (two mode model) has a similar structure as
our equation (\ref{18}), showing chaotic behavior, leading always
to a finite time for the period of the soliton trapping. This
phenomenon is due to stochastic instabilities inherent for
this dynamical system. The reason for this phenomenon is that the
finite dimensional system, like the one given by Eq.~(\ref{18}), cannot 
take into account the soliton radiation, that interacts with the defect. 
The effect of the radiation leads to the appearance of the damping in 
Eq.~(\ref{18}), that changes the long-time behavior of the system. In 
particular, the radiative damping can lead to the long-time regular 
dynamics~\cite{Good}.

For a more detailed investigation of the interaction between matter wave
soliton with the nonlinear impurity mode, excited on the inhomogeneity,
we need to develop collective coordinate approach like the one considered 
in Ref.~\cite{For}. This will be considered in future.

\section{Conclusion}
In this work, we have investigated the dynamics of bright matter 
wave soliton, in BEC systems with cigar type geometry and attractive 
interactions.
The inhomogeneities can appeared in BEC due to the existence of
regions in space with different values of the two-body atomic scattering
length $a_{s}$. These variations can be achieved using, for example, the
Feschbach resonances. Two kind of inhomogeneities in the spatial 
distribution of $a_s$ have been studied: local point-like and jump type.  
The first type has been modeled by a Dirac-delta 
function, that will result in a modulation of the nonlinear term in the 
GP equation, corresponding to the so called nonlinear impurity in the 
nonlinear Schr\"odinger equation. The second type corresponds to a 
sudden variation of the two-body scattering length, that affects 
the amplitude of the nonlinear term in GP equation. It correspond to 
the case that the BEC system is divided in two parts with different 
values of $a_s$ (see discussion of recent experiment in 
Ref.~\cite{Fatemi}).

The present investigation of the local variation in space of the atomic 
scattering length shows that different regimes of the soliton 
interaction with the nonlinear impurity are possible. It is observed
trapping, reflection and transmission regimes. The most interesting 
effect is the reflection of the atomic soliton by the attractive 
nonlinear impurity. We have also verified the occurrence of collapse of 
the soliton on the attractive impurity, when the strength of the 
impurity (or the initial number of atoms) exceeds a certain critical 
value.
This effect in quasi-1D BEC resembles the phenomena that occurs in 
2D BEC. Using the time-dependent variational approach we have described 
successfully both phenomena.

For the case of a sudden variation of the two-body scattering length, 
represented by a nonlinear jump inhomogeneity, using analogy with 
a nonlinear optical problem~\cite{Ace}, we presented the condition 
for the multiple bright matter soliton generation.

Finally, we would like to emphasize that the problem we have studied
in the present work has important application for the control of 
parameters of bright atomic matter solitons and for the generation of 
solitons in quasi 1D BEC.

\section*{Acknowledgments}
We are grateful to Funda\c{c}\~ao de Amparo \`a Pesquisa do
Estado de S\~ao Paulo (FAPESP) for partial financial support.
AG and LT also thank partial support from Conselho Nacional de
Desenvolvimento Cient\'\i fico e Tecnol\'ogico. Part of this work
was done by using resources of the LCCA - Laboratory of Advanced 
Scientific Computation of the University of São Paulo.

\end{multicols}
\end{document}